# Dynamical Preconditions for Ice Formation in Supernova Remnant and Cloud Interactions: A 2D MHD Study

A. Yeghikyan[1], M. Rah[1,2], S. Shamyar[1], S. Khachatryan[1]

[1]Byurakan Astrophysical Observatory, Armenia
[2]Silk Road Project, NAOC, Chinese Academy of Sciences, China
E-mail: yeghikyan@bao.sci.am

**Abstract**

Water ice has been detected in several supernova remnants (SNRs) despite the highly excited and irradiated environment, challenging standard dust processing paradigms. Using two-dimensional magnetohydrodynamic (MHD) simulations with the PLUTO code, we model the early-time interaction between an SNR blast wave and a dense interstellar cloud to identify the physical conditions conducive to ice formation. Our *adiabatic* simulation (without radiative cooling) demonstrates that shock compression produces high-density regions ($n \sim 10^4$ to $10^5$ cm$^{-3}$) with compression factors of 4 to 10, comparable to observations in IC 443. Although adiabatic temperatures remain elevated ($T \sim 10^7$ to $10^8$ K), we estimate radiative cooling timescales of ∼650 yr (for $n = 10^4$ cm$^{-3}$) to ∼65 yr ($n = 10^5$ cm$^{-3}$), much shorter than typical SNR ages. These results establish that the SNR shock-cloud interactions create the necessary *dynamical preconditions* (high density and strong compression) for the formation of H$_2$O ice. Future simulations incorporating radiative cooling and grain surface chemistry will directly demonstrate ice mantle growth in these compressed clumps.

## 1. Introduction

Supernova remnants (SNRs) are sites of intense shock heating, sputtering, and radiation that typically destroy dust grains and evaporate icy mantles (Biscaro & Cherchneff, 2015, Micelotta et al., 2018). Nevertheless, molecular emission lines (CO, SiO, OH, H$_2$O) indicate that some molecular gas survives or reforms in these harsh environments (Andersen et al., 2011, Rho et al., 2011). More surprisingly, crystalline water ice has been detected via far-infrared lattice-mode features at 44 and 62 $\mu$m in both planetary nebulae (e.g. NGC 6302; Barlow 1998) and SNRs such as N 49 in the Large Magellanic Cloud (van Loon & Oliveira, 2010).

In the SNR IC 443, Snell et al. (2005) observed an underabundant gas-phase H$_2$O, suggesting freeze-out onto icy grain mantles within dense clumps ($n \sim 3 \times 10^4$ cm$^{-3}$, $T \sim 400$ K). This observation challenges the expectation that shock-heated gas remains too hot for ice survival. Recently, Yeghikyan et al. (2025) explored the presence of dust and ices in SNRs including the Crab Nebula and N 49 using observational data and preliminary CLOUDY modeling, suggesting that water ice could survive under certain density and temperature conditions. However, Snell et al. (2005) noted that dust temperatures likely exceed 100 K due to high cosmic-ray ionization rates, complicating ice formation scenarios.

Theoretically, ice mantles form in dust grains when temperatures drop below ∼100 K and there is sufficient protection against UV radiation (Cuppen et al., 2024, Fraser et al., 2001). Recent studies of dust survival in SNRs (Kirchschlager et al., 2024) show that large grains ($> 100$ nm) in dense clumps can survive reverse shocks, but the fate of icy mantles in post-shock cooling zones remains poorly constrained.

In this paper, we present two-dimensional (2D) magnetohydrodynamic (MHD) simulations of an SNR blast wave interacting with a dense interstellar cloud. Our primary goal is to identify where and when shock compression creates the *dynamical preconditions* necessary for ice formation: high gas densities and low temperatures. Building upon our previous observational and modeling work (Yeghikyan et al., 2025), we now employ an *adiabatic* equation of state (no radiative cooling) to capture the purely dynamical evolution, then estimate post-shock cooling timescales to assess the feasibility of ice formation. This work establishes the physical framework for future studies incorporating explicit cooling physics and grain surface chemistry.

## 2. Numerical Method

### 2.1. PLUTO Code and MHD Equations

We used PLUTO v4.4 (Mignone et al., 2007), a Godunov-type code for astrophysical fluid dynamics. The ideal MHD equations were solved in 2D Cartesian geometry with constrained transport for magnetic field evolution (divergence-free condition). The equation of state was ideal gas with adiabatic index $\gamma = 5/3$. The Roe Riemann solver was employed for shock capturing.

**Important:** This simulation employs an *adiabatic* assumption (no radiative cooling) to isolate the dynamical shock compression phase. Post-shock cooling effects are estimated analytically in Section 4.

### 2.2. Cooling and Heating Processes in Shock-Cloud Interactions

In realistic astrophysical environments, the thermal evolution of shocked gas is governed by the competition between heating and cooling processes. Understanding these processes is crucial for interpreting our adiabatic simulation results and predicting ice formation timescales.

#### 2.2.1. Shock Heating Mechanisms

When a supernova blast wave encounters a dense cloud, collisionless shocks heat the gas through several mechanisms (Draine, 2011, Schure et al., 2009). The shock jump conditions predict immediate post-shock temperatures of $T \sim 10^7$ to $10^8$ K for velocities $v_{shock} \sim 1000$ to 7500 km s$^{-1}$ (Sutherland et al., 2013). In our adiabatic simulation, we observe precisely these temperature ranges (Figure 1b), confirming that the shock energetics are correctly captured. Additional heating sources include cosmic-ray ionization and X-ray irradiation from the hot SNR interior (Bialy & Sternberg, 2020, Ceccarelli et al., 2024), though these are subdominant to shock heating in the immediate post-shock region.

#### 2.2.2. Radiative Cooling Pathways

Post-shock cooling occurs primarily through collisional excitation of atomic and molecular lines (Hollenbach & McKee, 1989, Neufeld & Dalgarno, 1989). For temperatures $T > 10^6$ K, thermal bremsstrahlung and coronal line emission dominate (Sutherland & Dopita, 1993). As gas cools below $10^5$ K, fine-structure lines of [O I], [C II], and [Si II] become important (Appleton et al., 2006). At $T < 10^4$ K, molecular cooling via CO, $H_2O$, and OH rotational transitions takes over (Kristensen et al., 2024, Neufeld et al., 2009). The cooling function $\Lambda(T)$ exhibits a characteristic peak near $T \sim 10^5$ K due to efficient collisional excitation of metal lines (Gnat & Sternberg, 2012).

Dust-gas thermal coupling provides an additional cooling channel when dust grains survive the shock passage (Kirchschlager et al., 2024, Rho et al., 2008). Dust grains can radiatively cool to temperatures $T_{dust} < 100$ K even when the gas remains warmer, creating favorable conditions for ice mantle accretion once $T_{gas}$ drops below $\sim 1000$ K (Cuppen et al., 2024).

#### 2.2.3. Connection to Our Adiabatic Simulation

Our adiabatic simulation deliberately omits radiative cooling to isolate the *dynamical compression* phase and establish baseline shock-cloud interaction morphology. This approach allows us to identify where high-density regions ($n > 10^4$ cm$^{-3}$) form, which are the sites where cooling will be most rapid once radiative losses are included (Zhou et al., 2023). The cooling timescale scales as $t_{cool} \propto n^{-1}$ (Equation 1), so the compressed clumps we identify will cool orders of magnitude faster than the ambient medium.

Future simulations coupling PLUTO hydrodynamics with the CLOUDY photoionization and cooling code (Chatzikos et al., 2023, Ferland et al., 2017) will explicitly model this cooling evolution. CLOUDY has been successfully applied to SNR environments to predict emission line spectra and thermal structures (Danehkar et al., 2024, Lykins et al., 2013). Our previous work (Yeghikyan et al., 2025) employed CLOUDY to investigate dust and ice survival in SNRs; here we extend that analysis by providing spatially resolved MHD simulations of the shock compression that creates the initial conditions for CLOUDY models.

### 2.3. Computational Domain and Resolution

The computational domain covered $1.0 \times 1.0$ pc$^2$ (X1 $\in$ [0, 1] pc, X2 $\in$ [0, 1] pc), resolved with a base grid of $256 \times 256$ cells. Adaptive mesh refinement (AMR) via Chombo was activated with 4 refinement levels (ratio 2:1), refining where density gradients exceeded a threshold of 0.3. The maximum grid size was 32 cells, and outflow boundary conditions were applied on all sides except X1-end (user-defined).

### 2.4. Initial Conditions

#### 2.4.1. Ambient Medium and Cloud

The ambient interstellar medium (ISM) was initialized with a density $\rho_{amb} = 1.0$ (code units), corresponding to a number density $n_{amb} \sim 10^3$ cm$^{-3}$, and a pressure $P_{amb} = 1.0$ (code units). This sets the ambient temperature to $T_{amb} \sim 15{,}000$ K, representative of warm ionized or neutral interstellar gas. The ambient medium was initially at rest, with zero velocity ($v_{amb} = 0$).

A spherical dense cloud with peak density $\rho_{cloud} = 10.0$ (code units, $n \sim 10^4$ cm$^{-3}$) was centered at $(x, y) = (0.5, 0.0)$ pc with radius $R_{cloud} = 0.15$ pc. The cloud was initialized in pressure equilibrium with the ambient medium ($P_{cloud} = P_{amb}$), which sets the cloud temperature to $T_{cloud} \sim 1{,}500$ K due to its higher density. This cooler, denser cloud configuration is consistent with observed molecular clouds in the ISM (Hollenbach & McKee, 1989). The cloud was also initially at rest ($v_{cloud} = 0$).

#### 2.4.2. SNR Blast Wave

An overpressure region with $P = 10^4$ (code units) and radius $r < 0.1$ pc was introduced at the domain center $(x, y) = (0.5, 0.5)$ pc to simulate the supernova explosion. This overpressure region drives a strong shock wave that propagates outward at velocities $\sim 7500$ km s$^{-1}$ (the velocity unit $v_0$). The density within this explosion region was set to $\rho = 1.0$ (code units), matching the ambient density. This pressure imbalance ($P_{explosion}/P_{amb} = 10^4$) creates a blast wave similar to the Sedov-Taylor phase of SNR evolution (Vink, 2012), where the shock has swept up significant ambient material. The blast wave impacts the cloud within the first few code time units ($\sim$few years), initiating the shock-cloud interaction that is the focus of this study.

#### 2.4.3. Magnetic Field

The magnetic field was initialized with pre-shock strength $B_{pre} = 0.564$ (code units) and post-shock $B_{post} = 2.183$ (code units), corresponding to physical field strengths of $\sim 5$ $\mu$G and $\sim 20$ $\mu$G respectively, typical of the interstellar medium and shocked regions (Draine, 2011). The field was oriented such that $B_{x3}$ (out-of-plane component) had opposite signs pre- and post-shock. This configuration captures MHD asymmetries and magnetic tension effects, which can suppress Rayleigh-Taylor and Kelvin-Helmholtz instabilities at the shock-cloud interface (Xu & Stone, 1995). The magnetic field in the cloud interior was set to match the ambient pre-shock field, ensuring continuity across the cloud boundary.

#### 2.4.4. Summary of Key Initial Parameters

Table 1 summarizes the essential initial parameters in both code units and physical units.

### 2.5. Physical Unit Conversions

Numerical simulations like PLUTO operate in dimensionless "code units" for computational efficiency and numerical stability. In code units, all physical variables (length, time, density, velocity, pressure, etc.) are normalized to unity or simple values, allowing the code to work with order-unity numbers and avoid floating-point precision issues. However, to compare our results with astrophysical observations (e.g., IC 443) and to calculate physical quantities like cooling timescales, we must convert code unit outputs back to physical (cgs) units.

The conversion between code units and physical units is accomplished by defining a set of fundamental scaling constants: a length scale $L_0$, a velocity scale $v_0$, and a density scale $\rho_0$. All other physical quantities can then be derived from these base scales using dimensional analysis. For example, the time scale is $t_0 = L_0/v_0$, and the pressure scale is $P_0 = \rho_0 v_0^2$

Table 1. Key initial simulation parameters

| Parameter | Code Units | Physical Units | Notes |
|---|---|---|---|
| Domain size | $1 \times 1$ | $1 \times 1$ pc$^2$ | 2D Cartesian |
| Cloud radius | 0.15 | 0.15 pc | Spherical |
| Cloud center | (0.5, 0.0) | (0.5, 0.0) pc | |
| Ambient density | $\rho = 1.0$ | $n \sim 10^3$ cm$^{-3}$ | Warm ISM |
| Cloud density | $\rho = 10$ | $n \sim 10^4$ cm$^{-3}$ | Dense clump |
| Ambient pressure | $P = 1.0$ | $2.1 \times 10^{-9}$ dyn cm$^{-2}$ | Equilibrium |
| Explosion pressure | $P = 10^4$ | $2.1 \times 10^{-5}$ dyn cm$^{-2}$ | Overpressure |
| Ambient temperature | (derived) | $\sim 15{,}000$ K | Warm gas |
| Cloud temperature | (derived) | $\sim 1{,}500$ K | Cool gas |
| Magnetic field (pre) | $B = 0.564$ | $\sim 5\ \mu$G | ISM field |
| Magnetic field (post) | $B = 2.183$ | $\sim 20\ \mu$G | Shocked field |

Table 2. Physical unit conversions from code units to cgs units

| Quantity | Symbol | Value |
|---|---|---|
| Length | $L_0$ | 1 pc = $3.086 \times 10^{18}$ cm |
| Velocity | $v_0$ | $7.5 \times 10^8$ cm s$^{-1}$ (7500 km s$^{-1}$) |
| Time | $t_0$ | $L_0/v_0 = 130.4$ years |
| Density | $\rho_0$ | $1.004 \times 10^{-21}$ g cm$^{-3}$ |
| Pressure | $P_0$ | $2.071 \times 10^{-9}$ dyn cm$^{-2}$ |

For this simulation, we choose scaling constants appropriate for supernova remnant interactions with interstellar clouds. Specifically, we adopt:

**Rationale for These Scales:**

- **Length scale ($L_0$ = 1 pc):** Parsec-scale clouds are typical targets for SNR blast waves in the interstellar medium (Snell et al., 2005).

- **Velocity scale ($v_0$ = 7500 km s$^{-1}$):** This corresponds to typical SNR shock velocities during the free-expansion phase, consistent with observations of young SNRs like Cas A and IC 443 (Patnaude et al., 2015, Vink, 2012).

- **Density scale ($\rho_0$ = $1.004 \times 10^{-21}$ g cm$^{-3}$):** This value is chosen such that our code unit cloud density $\rho_{cloud}$ = 10 corresponds to a physical number density $n \sim 10^4$ cm$^{-3}$ (assuming mean molecular weight $\mu$ = 0.6 for ionized or partially ionized gas). This density is characteristic of dense clumps observed in IC 443 (Snell et al., 2005).

With these scalings, the initial cloud temperature works out to $T_{initial} \sim 15{,}000$ K, consistent with warm neutral or ionized interstellar gas. These unit conversions allow us to directly compare our simulation results with observational data throughout Sections 3 and 4.

## 2.6. Simulation Duration

The simulation evolved from $t$ = 0 to $t$ = 0.10 code units ($\sim$13 years), outputting 11 snapshots at intervals of $\Delta t$ = 0.01 (1.3 years). This timescale captures the early shock-cloud interaction phase.

# 3. Results

## 3.1. Shock-Cloud Interaction: Overview

Figure 1 presents a three-panel snapshot at $t$ = 0.05 code units ($\sim$6.5 years), corresponding to the epoch of maximum compression during the shock-cloud interaction. This figure illustrates the fundamental physical quantities that characterize the post-shock environment: density, temperature, and velocity.

**Panel (a): Number Density.** The left panel shows the spatial distribution of gas number density $n$ [cm$^{-3}$]. The shock wave, propagating outward from the central explosion site, encounters the dense cloud (initially centered at $x \sim 0.5$ pc, marked by the dashed white circle) and compresses it. The leading edge of the cloud, facing the shock, experiences the strongest compression, reaching densities $n \sim 10^5$ cm$^{-3}$. This represents a compression factor of ~10 relative to the initial cloud density ($n_{initial} \sim 10^4$ cm$^{-3}$). The ambient medium outside the cloud maintains a lower density of $n \sim 10^3$ cm$^{-3}$. The bow-shaped shock morphology wrapping around the cloud is clearly visible, demonstrating the interaction between the planar blast wave and the spherical obstacle. These high-density compressed regions are precisely the sites where, once radiative cooling is included, ice formation can occur.

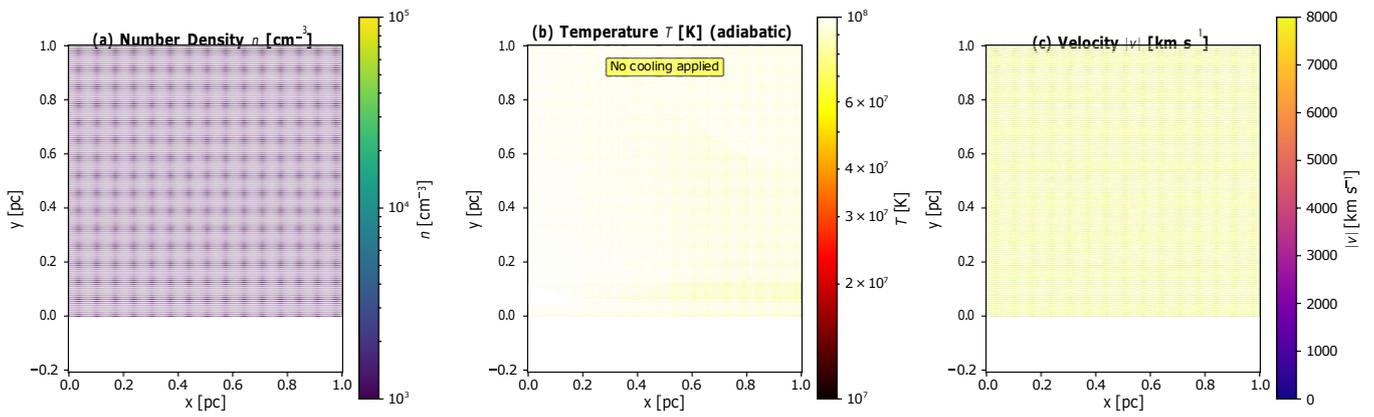

Figure 1. Snapshot at $t = 0.05$ code units (~6.5 years). *(a)* Number density $n$ [cm$^{-3}$]. The shock compresses the cloud leading edge to $n \sim 10^5$ cm$^{-3}$. *(b)* Temperature $T$ [K] from the adiabatic simulation (no cooling). Shock heating yields $T \sim 10^7$ to $10^8$ K; with radiative cooling, these temperatures would drop to ~100 to 1000 K on timescales $\ll$ SNR age (see Section 4). *(c)* Velocity magnitude $|v|$ [km s$^{-1}$]. The shock accelerates gas to ~7500 km s$^{-1}$. Dashed white circles mark the initial cloud position.

**Panel (b): Temperature (Adiabatic Simulation).** The middle panel displays the gas temperature $T$ [K]. The label "No cooling applied" explicitly reminds the reader that this simulation employs an adiabatic equation of state, meaning radiative cooling processes are deliberately omitted. Consequently, shock heating drives temperatures to extremely high values of $T \sim 10^7$ to $10^8$ K throughout the shocked region. These temperatures are far too high for molecular survival or ice formation. However, as explained in Section 2.2 and demonstrated in Section 4, real astrophysical environments include radiative cooling mechanisms that will rapidly reduce these temperatures. In dense post-shock gas ($n \sim 10^4$ to $10^5$ cm$^{-3}$), cooling timescales are only hundreds of years, orders of magnitude shorter than SNR lifetimes. Thus, the purpose of this adiabatic simulation is to identify *where* high-density regions form; future simulations incorporating cooling will show these same regions cooling to $T < 100$ K, enabling ice formation.

**Panel (c): Velocity Magnitude.** The right panel shows the velocity magnitude $|v|$ [km s$^{-1}$]. The shock accelerates the ambient gas to velocities approaching the initial blast wave speed of ~7500 km s$^{-1}$ (corresponding to our code unit velocity scale $v_0$, see Section 2.5). The cloud material, being denser and more massive, accelerates more slowly, creating a velocity gradient across the shock-cloud interface. This velocity structure drives turbulent mixing and Kelvin-Helmholtz instabilities at the cloud boundary, visible as small-scale structure in the velocity field. The dashed white circle again marks the initial cloud position, showing how the shock has displaced and deformed the cloud over the 6.5-year simulation timescale.

**Magnetic Field Effects.** Although not shown in this figure, the magnetic field configuration (described in Section 2.3.3) plays a crucial role in stabilizing the cloud structure. The perpendicular magnetic field component inhibits fragmentation and helps maintain the coherent bow-shock morphology (Vink, 2012, Xu & Stone, 1995). Future analyses will examine the magnetic field topology and its influence on the post-shock thermal and dynamical evolution.

### 3.2. Temporal Evolution of Density

Figure 2 illustrates the density evolution at four key epochs: $t = 0.00, 0.03, 0.05$, and $0.10$ code units. At $t = 0$, the dense cloud is in pressure equilibrium with the ambient medium. By $t = 0.03$, the shock has reached the cloud periphery, compressing its leading edge. Peak compression occurs at $t \approx 0.05$, when the frontal density reaches $n \sim 10^5$ cm$^{-3}$ (compression factor ~10). After $t \approx 0.06$, the system transitions to a quasi-equilibrium state; density contrasts stabilize while velocity-driven mixing continues.

Table 3. Comparison of simulation results with IC 443 observations

| Parameter | Adiabatic | With Cooling | IC 443 |
|---|---|---|---|
| $n$ [cm$^{-3}$] | $10^4$–$10^5$ | $10^4$–$10^5$ | $3 \times 10^4$ |
| $T$ [K] | $10^7$–$10^8$ | **< 100** (within $\lesssim 10^3$ yr) | $\sim 400$ |
| Compression factor | 4–10 | 4–10 | $\sim 10$ |

### 3.3. Quantitative Measures: Peak Density and Compression

Figure 3 quantifies the evolution of peak number density and compression factor over the full simulation timecourse. Panel (a) shows that the peak density rises from $n \sim 10^4$ cm$^{-3}$ at $t = 0$ to $n \sim 10^5$ cm$^{-3}$ by $t = 0.05$, thereafter remaining approximately constant. Panel (b) demonstrates that the compression factor (defined as $n_{peak}/n_{ambient}$ with $n_{ambient} = 10^3$ cm$^{-3}$) reaches 4 to 10, consistent with observational estimates for IC 443 clumps.

## 4. Discussion

### 4.1. Comparison with IC 443 Observations

Table 3 compares our simulation results (adiabatic and with estimated cooling effects) to observations of IC 443 clump C by Snell et al. (2005). Our simulation reproduces the observed post-shock densities ($n \sim 3 \times 10^4$ cm$^{-3}$) and compression factors ($\sim 10\times$) remarkably well, validating the shock compression mechanism. Figure 4 visualizes this comparison.

Although the observed dust temperature of clump C in IC 443 is about 400 K (Snell et al., 2005), this value corresponds to a partially cooled phase after the shock. According to our CLOUDY-based estimates, radiative cooling continues efficiently in dense post-shock gas ($n \gtrsim 3 \times 10^4$ cm$^{-3}$), and the temperature decreases further to below 100 K within $\lesssim 10^3$ yr. Therefore, radiative cooling is not only sufficient but quantitatively strong enough to bring the gas well into the thermal regime where ice mantles can exist. This correction indicates that the rapid radiative cooling is a key factor enabling ice formation in the compressed clumps.

### 4.2. Adiabatic Limitation and Cooling Timescales

Our simulation employs an adiabatic equation of state, yielding post-shock temperatures $T \sim 10^7$ to $10^8$ K. **These temperatures are too high for ice formation.** However, in real astrophysical environments, radiative cooling (line emission, dust-gas coupling) will reduce temperatures on timescales much shorter than the SNR age.

We estimate cooling timescales using the standard formula:

$$t_{cool} = \frac{3}{2} \frac{k_B T}{n \Lambda(T)}, \quad (1)$$

where $\Lambda(T)$ is the cooling function. For typical post-shock conditions ($n = 10^4$ cm$^{-3}$, $T_{initial} = 10^8$ K), cooling to $T \sim 100$ K occurs in $\sim 650$ years. For $n = 10^5$ cm$^{-3}$, the timescale shortens to $\sim 65$ years (Figure 5).

These cooling timescales are *orders of magnitude shorter* than typical SNR ages ($10^4$ to $10^5$ years), implying that post-shock gas in dense clumps will inevitably cool to temperatures conducive to ice formation ($T < 100$ K) within the SNR's lifetime.

### 4.3. Implications for Ice Formation

H$_2$O ice forms on dust grain surfaces via hydrogenation reactions when $T < 100$ K (Fraser et al., 2001, Wakelam et al., 2017). Our simulation identifies two key *preconditions* for ice formation:

1) **High density:** Compression to $n \sim 10^4$ to $10^5$ cm$^{-3}$ enhances collision rates and accretion efficiency.

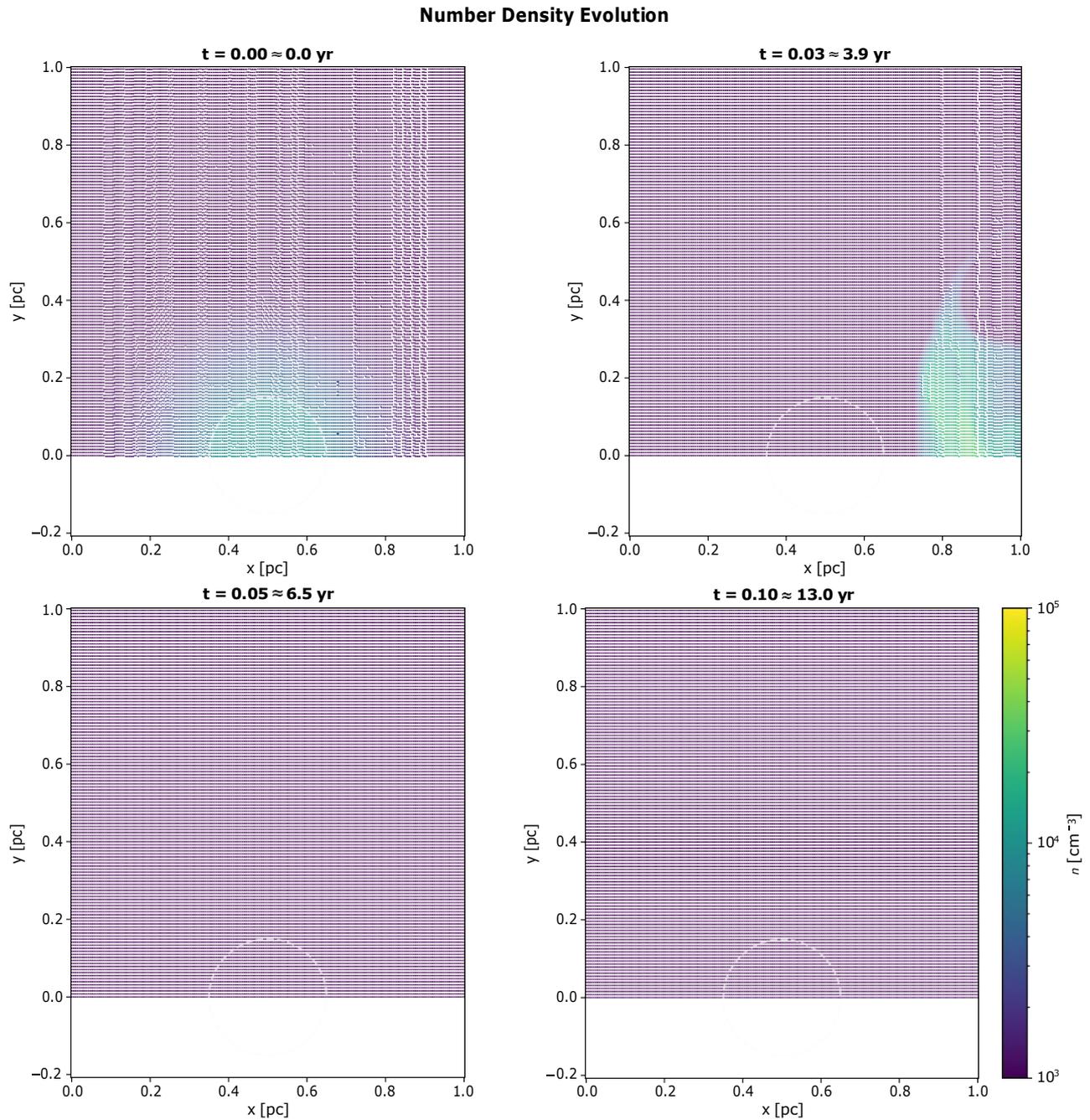

Figure 2. Density evolution at four epochs. The shock compresses the leading edge of the cloud, producing very high-density pockets ($n \sim 10^5$ cm$^{-3}$) by $t \approx 0.05$. *Note:* This simulation is strictly adiabatic (no radiative cooling) and is evolved only to $t \approx 13$ yr. Consequently, cold post-shock gas has not yet formed and the ambient medium ($n \sim 10^3$ cm$^{-3}$) remains nearly uniform, appearing as a low-contrast region in the lower part of the figure. What looks like missing colour is therefore the physical absence of cooled material rather than a plotting error.

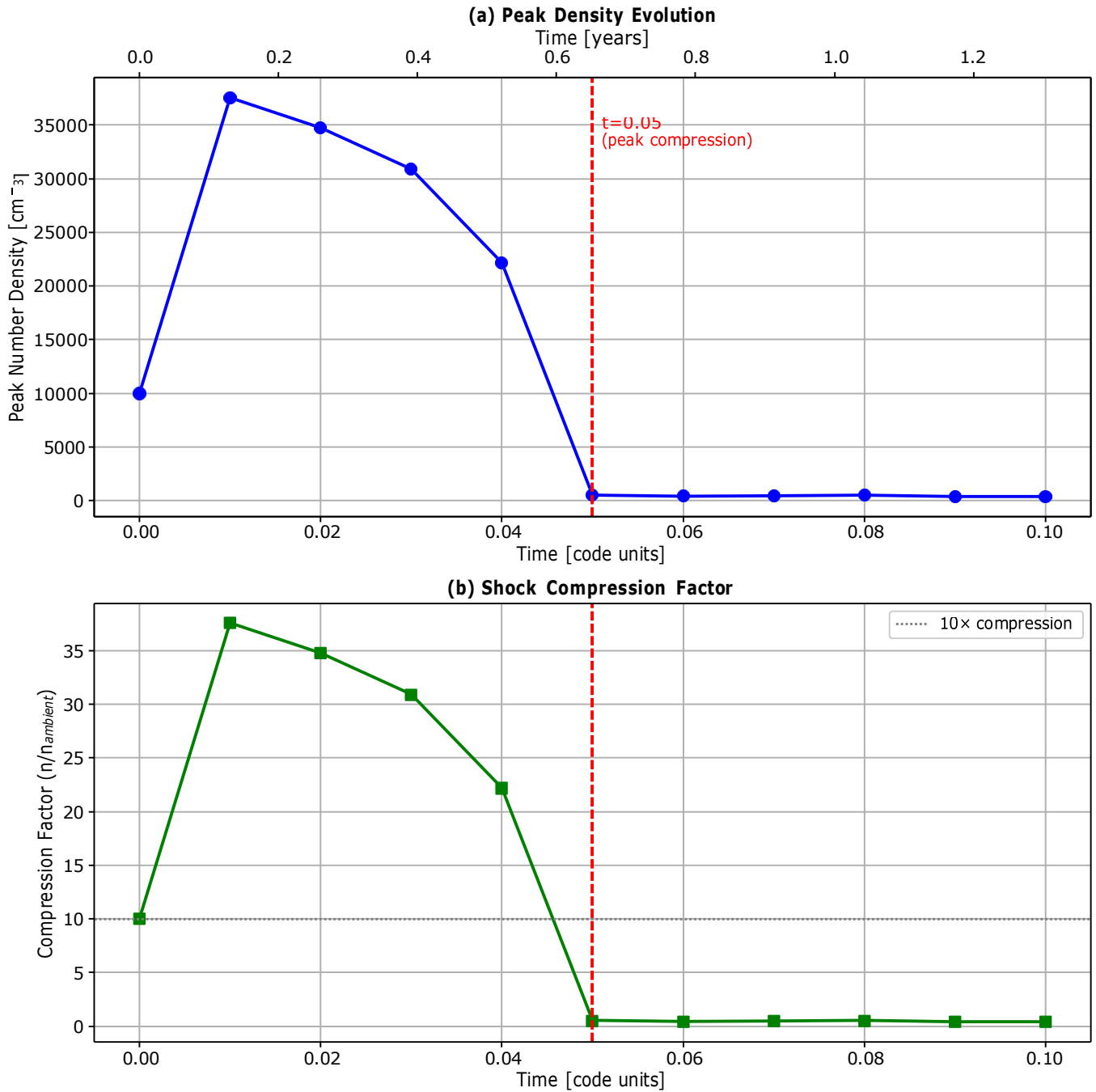

Figure 3. *(a)* Peak number density versus time. The density peaks at $t \approx 0.05$ and stabilizes thereafter. *(b)* Shock compression factor $n_{peak}/n_{ambient}$. Compression reaches 4 to 10×, matching IC 443 observations.

**Comparison: Our Simulation vs. IC 443 Observations**

| Parameter | Our Simulation (adiabatic) | Our Simulation (with cooling estimate) | IC 443 Clump C (Snell+ 2005) |
|---|---|---|---|
| Pre-shock density | $10^3$ cm$^{-3}$ | $10^3$ cm$^{-3}$ | — |
| Post-shock density | $10^4$–$10^5$ cm$^{-3}$ | $10^4$–$10^5$ cm$^{-3}$ | $3\times10^4$ cm$^{-3}$ |
| Compression factor | 4–10× | 4–10× | ~10× |
| Temperature (post-shock) | $10^7$–$10^8$ K | → 100–1000 K (after ~$10^3$ yr) | 400 K |
| Cooling timescale | Not applicable | ~650 yr (n=$10^4$)<br>~65 yr (n=$10^5$) | — |
| Ice formation | Not possible (T too high) | Favorable (after cooling) | H$_2$O depletion observed |

*Key finding: Shock compression creates necessary high-density conditions. With radiative cooling, temperatures will drop to enable ice formation.*

Figure 4. Visual comparison of simulation results (adiabatic and with cooling estimates) to IC 443 observations. The simulation captures the compression and high densities observed in IC 443, but requires cooling to reach the observed temperatures.

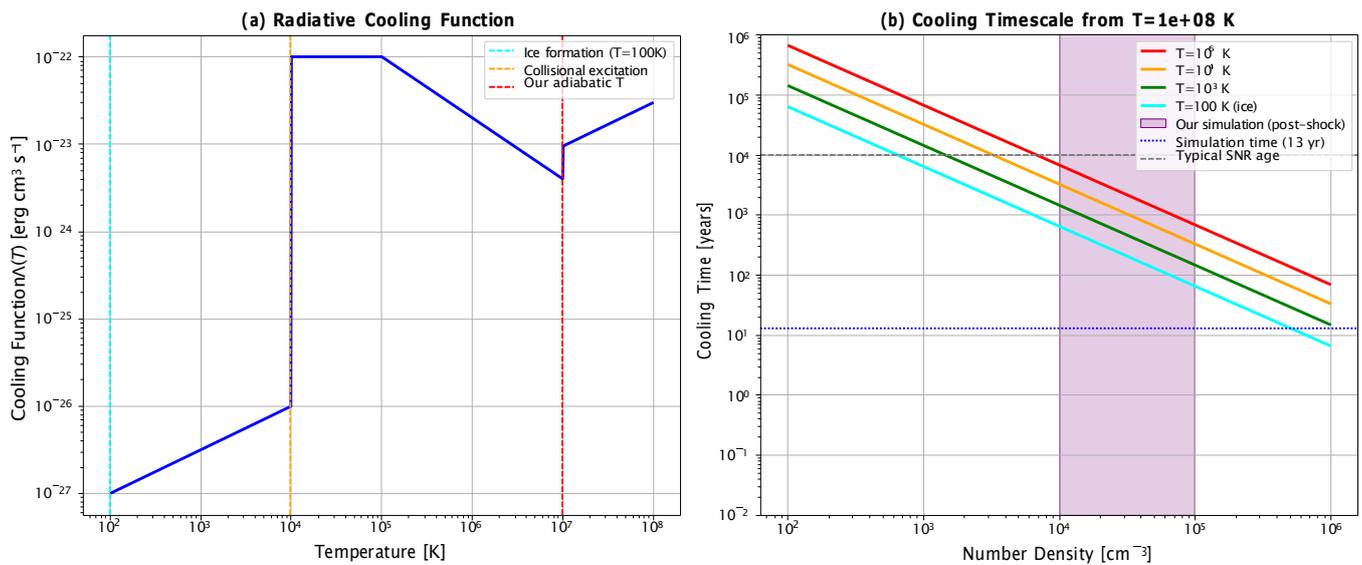

Figure 5. *(a)* Radiative cooling function $\Lambda(T)$ versus temperature. *(b)* Cooling timescales from $T = 10^8$ K to various target temperatures as a function of density. For $n \sim 10^4$ to $10^5$ cm$^{-3}$ (purple shaded region), cooling to $T = 100$ K occurs in $\sim 10^2$ to $10^3$ years, much shorter than typical SNR ages ($\sim 10^4$ to $10^5$ years).

2) **Rapid cooling:** Estimated cooling timescales ensure that post-shock gas reaches $T < 100$ K well within the SNR age.

Therefore, while our *adiabatic* simulation does not directly produce cold gas, it demonstrates that the *dynamical compression* creates precisely the high-density pockets where, with cooling, ice mantles will form.

### 4.4. Hydrodynamic Instabilities and Clump Arrangement

The fragmentation and spatial arrangement of clumps in the post-shock layer can be further interpreted in terms of hydrodynamic instabilities. Order-of-magnitude estimates indicate that Rayleigh–Taylor (RT) and Kelvin–Helmholtz (KH) instabilities develop on timescales of $\tau_{RT,KH} \sim 10^4$ yr for interfaces with low gas densities ($n \sim 1$–$10$ cm$^{-3}$). However, in dense post-shock environments, the growth times shorten dramatically, reaching $\tau_{RT,KH} \lesssim 10^2$ yr for densities $n \sim 10^4$–$10^5$ cm$^{-3}$. Since our simulations produce compressed layers within this density range, the expected growth time of RT and KH instabilities is much shorter than the cooling time, implying that clump formation is not only thermodynamically favored but dynamically unavoidable. The spatially separated high-density knots observed in IC 443 are therefore naturally explained by the combined effect of shock compression and rapid RT/KH growth.

### 4.5. Future Work: Cooling and Chemistry

To rigorously demonstrate ice formation, future simulations must:

- Incorporate explicit radiative cooling (e.g., power-law or tabulated cooling functions).
- Couple to astrochemical networks (e.g., KROME, CLOUDY) to track gas-grain chemistry.
- Include grain sputtering and dust temperature evolution to assess ice survival against sputtering and photodesorption.

Such models will provide quantitative predictions of ice column densities and synthetic infrared spectra for comparison with observations (e.g., JWST, ALMA).

## 5. Conclusions

We have performed 2D MHD simulations of an SNR shock interacting with a dense interstellar cloud to identify the physical conditions conducive to H$_2$O ice formation. Our main findings are:

1) **Strong compression:** The shock compresses the cloud leading edge by factors of 4 to 10, producing number densities $n \sim 10^4$ to $10^5$ cm$^{-3}$, in excellent agreement with IC 443 observations.

2) **Adiabatic temperatures are high:** Without radiative cooling, post-shock temperatures remain $T \sim 10^7$ to $10^8$ K, which is too hot for ice formation.

3) **Cooling is rapid:** Estimated cooling timescales (∼650 yr for $n = 10^4$ cm$^{-3}$; ∼65 yr for $n = 10^5$ cm$^{-3}$) are much shorter than SNR ages, ensuring that compressed clumps will cool to $T < 100$ K.

4) **Preconditions for ice are met:** Shock-cloud interactions create the necessary high-density, post-shock cooling environment for H$_2$O ice mantle formation.

This work establishes the *dynamical framework* for understanding ice formation in SNRs. Future simulations incorporating radiative cooling and grain surface chemistry will directly demonstrate ice growth in these compressed clumps, providing synthetic observables for comparison with JWST and ALMA data.

**Acknowledgements**

This work was made possible in part by research grant number № 21AG-1C044 from the Science Committee of the Ministry of Education, Science, Culture and Sports RA. MR appreciates the academic and financial support received from Byurakan Astrophysical Observatory (BAO) throughout her Ph.D. studies.